\documentclass[epj]{svjour}

\usepackage{graphicx}
\usepackage{fancyhdr}
\def\makeheadbox{{
 }}
\def\mytitle{My title} 
\def\myauthors{My name}  
\def\mytype{My type of session}
\def\mysession{My session}

\def\mytitle{Trigger Strategies for SUSY Searches at the LHC} 
\def\myauthors{A.~De Santo}    
\def\mytype{Contributed Talk}    
\def\mysession{Colliders - SUSY Phenomenology}

\begin{document}
\title{Trigger Strategies for SUSY Searches at the LHC}
\author{Antonella De Santo (for the ATLAS and the CMS collaborations)\inst{}
\thanks{\emph{Email:}antonella.de-santo@rhul.ac.uk}%
}                     
%
%
\institute{Royal Holloway, University of London, Egham Hill, Egham, Surrey, TW20 0EX, UK
}
%
\date{}
\abstract{
Supersymmetry will be searched for in a variety of final states at the LHC. It is 
crucial that a robust, efficient and unbiased trigger selection for SUSY is 
implemented from the very early days of data taking. After a brief description of 
the ATLAS and the CMS trigger systems, and a more in-depth discussion of the ATLAS 
High-Level Trigger, a triggering strategy is outlined for early SUSY searches at the 
LHC.
\PACS{
      {PACS-key}{ATLAS}  \and
       {PACS-key}{CMS}     \and
       {PACS-key}{SUSY}     \and
       {PACS-key}{High-Level Trigger}
     } 
} 
\maketitle
\section{Introduction}
\label{intro}
With the LHC start up next year,
the largely unexplored domain of
multi-TeV scale physics will finally become accessible experimentally 
at a collider.
It is widely believed that both ATLAS and CMS will observe new physics
beyond the Standard Model (SM), and supersymmetry (SUSY) is certainly one of 
the theoretically favoured SM extensions.
Appealing features of SUSY are the fact that it 
provides natural cancellations 
to the higher mass corrections to the Higgs mass, that it unifies the
electroweak and the strong force at the GUT scale, and that it 
provides a good dark matter candidate.

The SUSY cross-section at LHC energies is dominated by
gluino and squark pair production, whose decay will typically give
rise to multi-jet, high-pt final states. 
Moreover, in R-parity conserving
models, due to the escaping LSP (Lightest Supersymmetric Particle), 
SUSY states will be characterized by
a large energy imbalance in the plane transverse to the beam direction
(large ``missing ET'', or MET).
Often one or more isolated leptons, from the decay of intermediate
particles in the decay chain, will also be present in the final state.

Rare new physics processes, including SUSY ones, will have to be discriminated 
against a very large background of SM
events. A sophisticated online system is hence required to apply fast
and reliable
signature-based selection algorithms, which must in turn deliver
the required efficiency without introducing significant biases to the
data written ``on tape''.

ATLAS\cite{ref:ATLAS_general} and CMS\cite{ref:CMS_general} have both developed highly complex trigger
systems which, despite the significant differences in their
architectures, perform rather similarly and give comparable output rates 
and efficiencies. In the following, the underlying design of both systems,
with particular reference to the ATLAS High-Level Trigger (HLT), is
discussed, and a strategy is described 
to select SUSY events online. Preliminary simulation-based
results for relevant trigger menus
at ``initial'' luminosity ($L=10^{31-32}~\textrm{cm}^{-2}~\textrm{s}^{-1}$)
are also given.

\begin{figure}[h]
\includegraphics[width=0.45\textwidth,height=0.65\textwidth,angle=0]{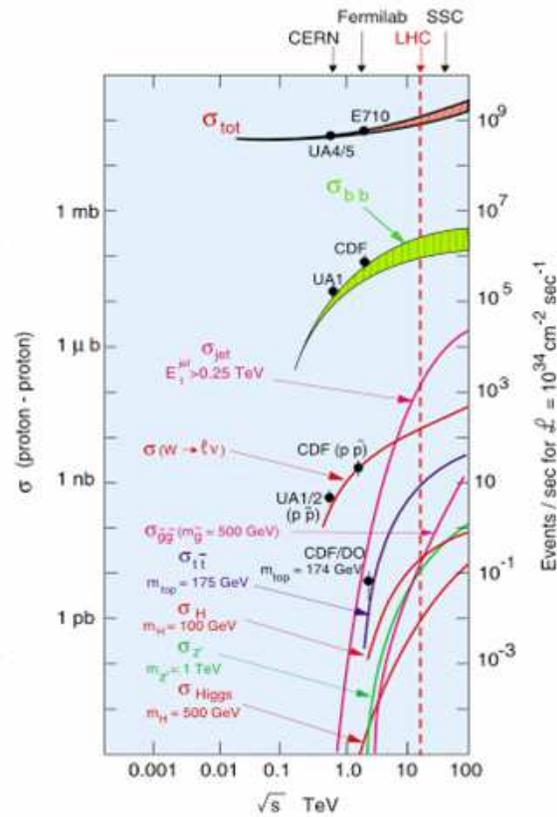}
\caption{Cross-sections vs. centre-of-mass energy for proton-proton interactions. Rates at 
$L=10^{34}~\textrm{cm}^{-2}~\textrm{s}^{-1}$ are also given.}
\label{fig:cross_sections}
\end{figure}

\begin{figure*}[ht]
\begin{tabular}{cc}
\includegraphics[width=0.45\textwidth,height=0.34\textwidth,angle=0]{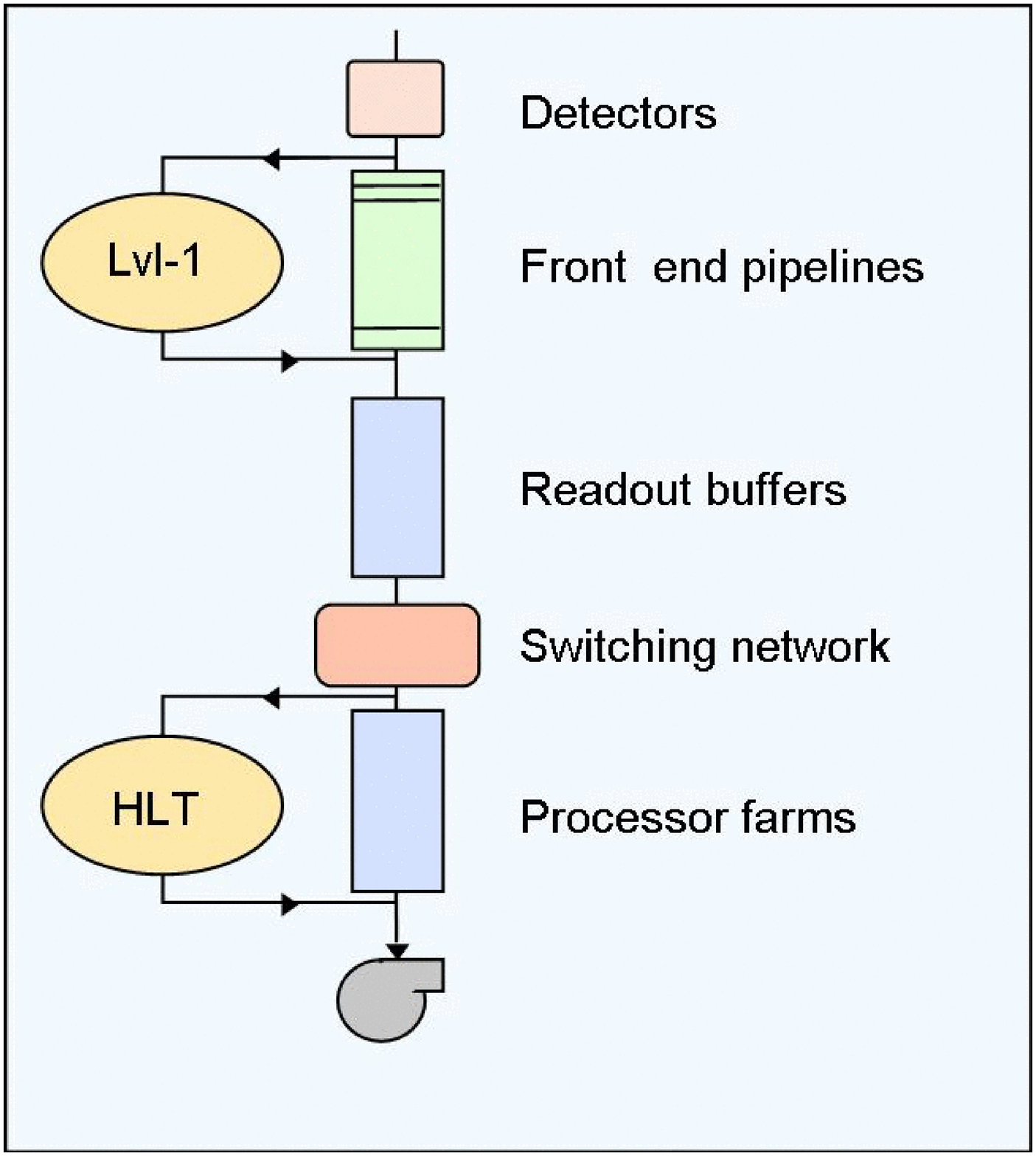}&
\includegraphics[width=0.45\textwidth,height=0.34\textwidth,angle=0]{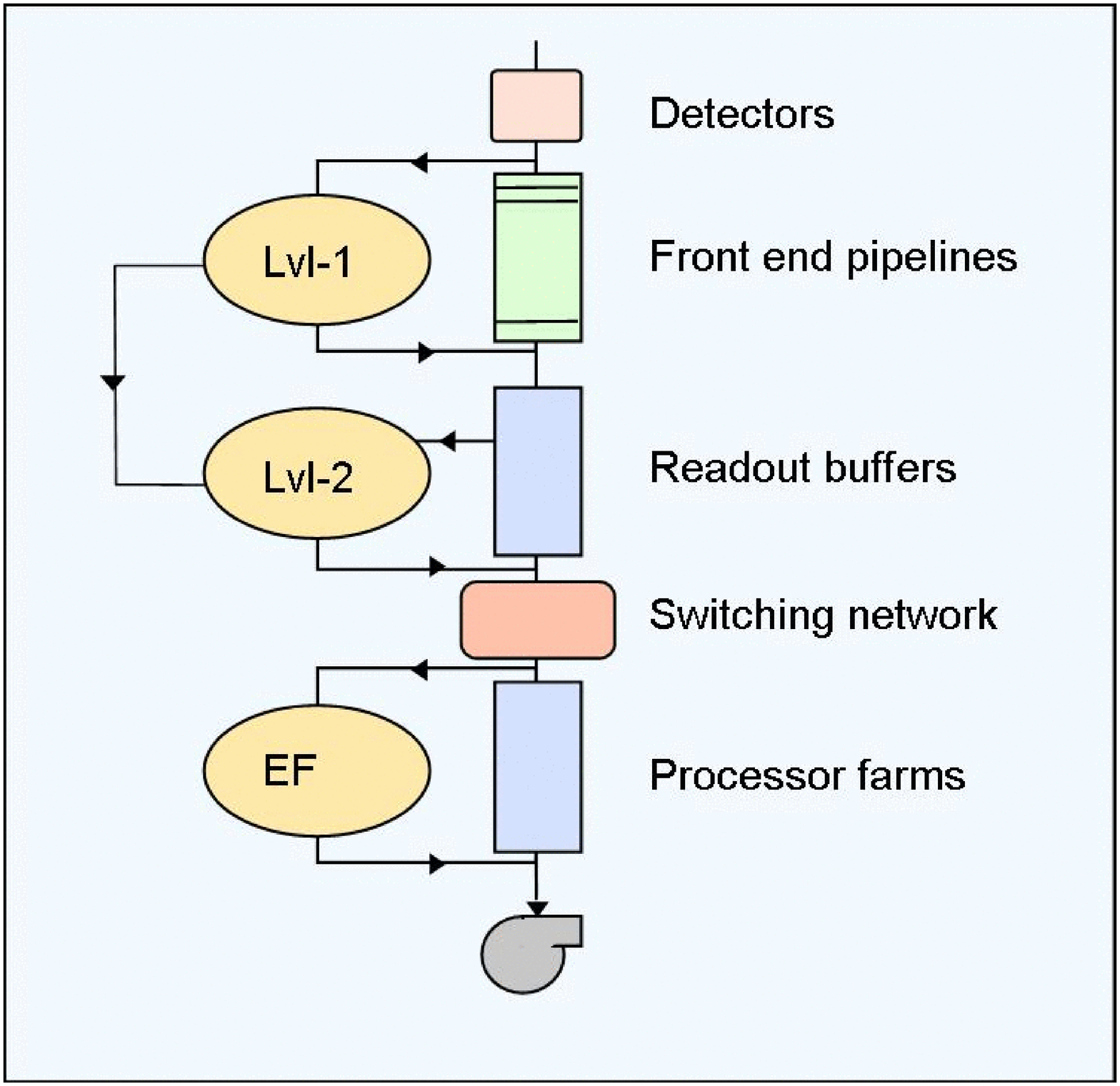}\\
\end{tabular}
\hfill
\caption{Schematics of the CMS (left) and the ATLAS (right) trigger
  architectures.}
\label{fig:trigger_design} 
\end{figure*}

\section{The ATLAS and the CMS Trigger Systems}\label{sec:trig_sys}

Typical cross-sections for different processes in proton-proton
interactions, as well as the corresponding
event rates at nominal ``high''luminosity
($L=10^{34}\textrm{cm}^{-2}\textrm{s}^{-1}$), 
are shown in  Fig.\ref{fig:cross_sections} 
as a function of the
centre-of-mass energy. The total cross-section at 14~TeV
(O($110$~mb)) is largely dominated by
soft inelastic $pp$ interactions (``minimum bias'', or MB; $\sigma_{MB}\sim
80~\textrm{mb}$), which give rise to a high-rate
(O($10^9$~Hz) at high L)
of events with low-$p_T$ particles
in the final state. At high luminosity, on average 23 MB events will be
superimposed to any interesting high-$p_T$ interaction, such as for
example leptonic decays of
the Z or the W bosons. However, while it will be relatively easy to
separate between MB events and generic
hard scatters, enriching the high-$p_T$ sample online with rare
processes such as gluino pair production, or even leptonic decays of
the Higgs, will not be as straightforward. To achieve that, 
both ATLAS and CMS will rely on
their trigger system's capability to apply fast algorithms to signals
from their calorimeters and muon detectors, and to select events with
leptons, jets and large MET in the final state.

The LHC $25$~ns bunch crossing determines the trigger input rate of
$40$~MHz, which has to be reduced to a more manageable $\sim 100$~Hz
of interesting physics events to be kept on permanent storage. With an
average event size of $1$-$2$~Mbytes, this corresponds to about
$1$-$2$~PByte worth of data to be recorded each year.

In both experiments the required selection capabilities of the trigger are
achieved via consecutive decision stages. 
A schematic representation of the trigger architectures for the
two experiments is given in Fig.\ref{fig:trigger_design}. Both ATLAS
and CMS have a hardware-based ``Level-1'' (Lvl-1) trigger, while further
selections are performed at software level 
in the successive stage of the so-called High-Level Trigger (HLT). In
the case of CMS, this 
consists of just one extra trigger level, while in the case of ATLAS
the HLT is further subdivided into ``Level-2'' (Lvl-2) and ``Event
Filter'' (EF).

In the hardware-based Lvl-1
trigger of both ATLAS and CMS, high-speed
pipelined front-end electronics, custom-made for the experiments, 
gets access to coarse granularity information
from the calorimeters and the muon system, and then
runs simple selection algorithms to
make decisions about the events. This is done 
synchronously with the machine bunch
crossing, with a latency time of $2.5~\mu$s and 
$3.2~\mu$s for ATLAS and CMS respectively.
The reduction factor that the Lvl-1 must
achieve within this time is of $\sim 400$ for both experiments. This
brings the input rate for the second stage of the trigger down to 
$\sim 100$~kHz.

The HLT, which in both cases is implemented using farms of fast
commercial processors running sophisticated reconstruction
algorithms, has a very different architecture for ATLAS and CMS. While
in CMS the full event is available at the HLT input, the ATLAS trigger
is so designed that on average the Lvl-2 only needs to access 
a fraction ($<10\%$) 
of the total event. This greatly reduces the
requirements on the available bandwidth for ATLAS 
at that stage of the selection.
A description of the ATLAS HLT is given below, while details of the CMS
HLT are discussed elsewhere in these conference proceedings~\cite{ref:cms_hlt_susy07}.

\subsection{The ATLAS High-Level Trigger}\label{sec:ATLAS_HLT}

In ATLAS, the Lvl-2 and EF trigger levels are collectively known as the
HLT. The overall necessary $\sim 10^3$ reduction factor in the rate, 
from the O($100$~kHz) at the Lvl-2 input to
the final O($100$~Hz) at the EF output, is achieved in two
steps. A first suppression factor
($\sim 100$) is provided by the Lvl-2, bringing the EF input rate down to 
O($1$~kHz), while the extra factor of $\sim 10$ comes from the
EF itself. For an $8$~GHz processor, 
the Lvl-2 and EF processing times are $\sim 10$~ms and
$\sim 1$~s respectively. 

The way ATLAS can achieve the required trigger 
performance by using a smaller bandwidth
than CMS is by implementing the concept of the so-called
``Regions-of-Interest'' (RoI), which are built at Lvl-1 and then passed
on to Lvl-2 for further analysis, if the event 
survives 
the Lvl-1 selection. If the event also passes the Lvl-2 selection,
the full event is then built and transferred to the EF for processing.

The basic idea is to use a ``seeded'' and
``stepwise'' selection strategy, which makes it possible to accomplish
early rejection of
uninteresting events with minimal amount of processing. 
In practice, based on coarse detector
granularity, the Lvl-1 processor constructs
objects in the calorimeters and in the muon system which
are then analysed
under a specific trigger hypothesis. For
example, the properties of a cluster in the electromagnetic
calorimeter would be tested to verify whether or not they
are compatible with those of an electromagnetic shower. If they are,
the coarsely reconstructed cluster will ``seed'' the building of an RoI, 
which is then passed on to
the following trigger level.
At Lvl-2, full detector granularity as well as tracking information
are accessible within the RoI. This, in the example above, would allow 
the Lvl-2 algorithms to check
whether an inner detector track could be matched to the seed 
cluster found at Lvl-1.
As a consequence, discrimination between electrons and
photons becomes possible at Lvl-2. Events surviving the Lvl-2
selection criteria are then passed on to the EF, 
which also has access to tracking information and to full detector
granularity, now for the full event. As the EF can use significantly 
more time than the Lvl-2 to analyze each event, it can also use slower
but more refined
reconstruction algorithms. Therefore, while
dedicated fast ``online'' algorithms are run at Lvl-2, the EF uses a 
reconstruction that is close to that used for 
``offline'' analysis. Based on the outcome of the EF selection,
a final decision is made as to whether to transfer the current
event on to permanent storage, or whether to discard it irreversibly instead.

From the above it is clear that the ATLAS 
HLT is a highly flexible system, where it is relatively
straightforward to combine 
single trigger hypotheses into more complex trigger menus needed for
physics analysis. This feature is particularly
useful in the case of SUSY triggers, for which it will be possible to
rely on a very rich phenomenology to devise a redundant and efficient
selection strategy.

\section{SUSY Triggers in ATLAS and CMS}\label{sec:susy_strategy}

Because it is not possible to anticipate which specific supersymmetric
model
is actually realised in Nature, if any, the overall strategy for
SUSY searches
at the LHC will have to be as generic as technically feasible,
encompassing the maximum number of 
experimentally observable signatures. 
This is even more true for the trigger than it is for the offline
analysis, as events lost at trigger level are lost forever. No later
improvements in the selection techniques or in the reconstruction 
can help recover them, if they have
not been permanently saved to storage at the online stage.

A variety of studies have been developed in both ATLAS and CMS
to understand the triggering issues of more ``exotic'' SUSY
signatures, such as non-pointing photons in GMSB models, or highly
ionizing muon-like signals from R-hadrons. Some of these aspects have
been discussed elsewhere in these conference
proceedings~\cite{ref:r-hadrons},~\cite{ref:gmsb}. In this paper,
however, the focus will be on the more typical SUSY signatures from
mSUGRA motivated R-parity conserving models: high-$p_T$ multi-jets, large MET,
and possibly one or more leptons in the final state.

It is very important to realize that, among the 
classic SUSY signatures mentioned above, some 
will be more robust than others. For the very
early stages of data taking, particular care will have to be
paid not to rely too heavily on quantities that may take significant
time to be correctly understood, and may therefore initially introduce
significant biases in the data. 
For example, experience from past and current
experiments at hadron colliders suggests that the reconstruction of
the MET variable takes significantly more time than others to become
established. Several instrumental effects
can contribute to overestimate the high-end tail of the MET
distribution, which is where a SUSY signal would typically
be expected to be observed. Moreover, as MET is a very good variable
to discriminate
between SUSY events and SM backgrounds, it is
crucial that unbiased samples are collected, where MET can be used in
the offline analysis to define the boundaries between 
control and signal regions. All of the above strongly suggests the need to
de-emphasize the use of MET triggers in the early days of data
taking, when the low luminosity will allow low-threshold jet triggers
with affordable rates. For the same reasons, also at 
$L=10^{33}\textrm{cm}^{-2}\textrm{s}^{-1}$ it will be important to
keep the MET threshold as low as possible. Similarly, and 
again for the sake of systematic studies at the analysis phase,
too tight criteria for the
selection of leptons should also be avoided at trigger level.

The LHC will not reach its design luminosity for quite some time after
the start up, and even ``low'' values of $L$ (in the range of 
$10^{33}\textrm{cm}^{-2}\textrm{s}^{-1}$) will not be accessible in the
early stages of the data taking. For this reason both ATLAS and CMS
are developing lists of trigger menus especially conceived 
for ``early data'' scenarios,
assuming ``initial''
benchmark luminosities of
$10^{31-32}\textrm{cm}^{-2}\textrm{s}^{-1}$. Although
the list of
available triggers is likely to change over time, to adapt to the changing
experimental conditions, and potentially 
to cope with higher-than-expected trigger rates,
it is very important that well defined triggering strategies
are in place ahead of the start of the data-taking. This will 
ensure that interesting events can
be selected efficiently without exceeding the total rate budget available
at each trigger level.

\begin{table}
\caption{CMS HLT rates at $L=10^{32}\textrm{cm}^{-2}\textrm{s}^{-1}$
  for some of the trigger menus relevant for SUSY
  searches~\cite{ref:cms_hlt}. The
  numbers in the second column give the $p_T$ thresholds for the
  corresponding object in the trigger menu. The total CMS HLT expected
  output
  at this luminosity is $\sim 150$~Hz.}
\label{tab:cms_rates}
\begin{center}
\begin{tabular}{ccc}
\hline\noalign{\smallskip}
HLT path & $p_T$ and MET      & Rate (Hz) \\
         & threshold(s) (GeV) &           \\
\noalign{\smallskip}\hline\hline\noalign{\smallskip}
$1$-jet & $200$ & $9.3\pm 0.1$  \\
$2$-jets & $150$ & $10.6\pm 0.0$  \\
$3$-jets & $85$  & $7.5\pm 0.1$  \\
$4$-jets & $60$  & $3.9\pm 0.1$  \\
\noalign{\smallskip}\hline\noalign{\smallskip}
MET     & $65$  & $4.9\pm 0.7$  \\
$1$-jet$+$MET     & $(180,60)$  & $2.2\pm 0.1$  \\
$2$-jets$+$MET     & $(125,60)$  & $1.0\pm 0.0$  \\
$3$-jets$+$MET     & $(60,60)$   & $0.6\pm 0.0$  \\
$4$-jets$+$MET     & $(35,60)$   & $1.2\pm 0.1$  \\
\noalign{\smallskip}\hline\noalign{\smallskip}
$e$+jet           & $(12,40)$   & $11.6\pm 1.2$ \\
$\mu$+jet         & $(7,40)$    & $6.3 \pm 0.7$ \\
\noalign{\smallskip}\hline\hline\noalign{\smallskip}
\end{tabular}
\end{center}
\end{table}

As an example, a list of trigger rates is given 
in Tab.\ref{tab:cms_rates}~\cite{ref:cms_hlt} 
for a choice of HLT trigger paths from
the CMS experiment, for an assumed initial luminosity of 
$10^{32}\textrm{cm}^{-2}\textrm{s}^{-1}$.
All trigger signatures in the table are 
relevant for SUSY searches.
As it is apparent from these rate figures, provided
that sufficiently high thresholds are used to apply cuts on 
particle $p_T$ values and on global event variables such as MET, 
acceptably low rates (O($10$ Hz)
or less) can be achieved for each trigger menu.

ATLAS uses a very similar trigger strategy for early physics
running. Trigger menus are in place for a luminosity of
$10^{31}\textrm{cm}^{-2}\textrm{s}^{-1}$ and new ones are being
developed for $10^{32}\textrm{cm}^{-2}\textrm{s}^{-1}$.
In the current ATLAS framework,
initial 
trigger selections of SUSY events are mos-tly based on Lvl-1 menus. These
achieve adequately low trigger rates, provided that appropriate
$p_T$ and energy thresholds are chosen for the selection.
For example, at $10^{31}\textrm{cm}^{-2}\textrm{s}^{-1}$,
using the same notation as in Tab.\ref{tab:cms_rates},  
single trigger menus like ``$1$-jet'' or ``$e$+MET'' both have
rates of $\sim 10$~Hz, for $p_T$ and $(p_T$,MET$)$ 
thresholds of $100$~GeV and $(20,15)$~GeV
respectively.

In ATLAS, to evaluate 
the trigger performance on SUSY events, the trigger
efficiency has been calculated at each step for a number of
offline SUSY analyses, and in particular those for  
the inclusive channels. The trigger efficiency has been normalized at
each stage to the number of events surviving the SUSY selection at
that level. Typically, for the
$10^{31}\textrm{cm}^{-2}\textrm{s}^{-1}$ menus, 
standard jet triggers achieve
efficiencies very close to $95-100$\% in a large fraction of the
cosmologically relevant part of the mSUGRA parameter space, with the
efficiency being higher for high-jet multiplicities and 
at higher-mass SUSY points.
The effect of jet trigger rate uncertainties has also been studied in ATLAS.
If the jet rates were significantly higher than expected, the jet
$p_T$ thresholds would have to be increased
considerably to keep the rates at an acceptable level. Preliminary
studies show that, should that be the case, SUSY events would still be
selected with high efficiency.

\section{Conclusions}
Supersymmetry is one of the most appealing models of new physics to be
searched for at the LHC. If SUSY exists, and it is 
accessible at LHC energies, the richness of its phenomenology
can be used to devise redundant trigger menus that can be used to  
extract a significant supersymmetric 
signal from the dominant SM background.
The performance of the  
trigger will be a crucial element
of the analysis flow, and it is essential that the system is capable 
to select SUSY events in a manner that is both efficient and bias-free.
ATLAS and CMS both have very sophisticated trigger systems, which have
been shown to perform adequately for the stated purpose.

%

%
%

\end{document}